\documentclass[prb,aps,twocolumn,preprintnumbers,10pt,floats,amsmath,showpacs,amssymb,superscriptaddress]{revtex4-1}
\usepackage{epsfig}
\usepackage{amsmath}
\usepackage{amssymb}
\usepackage{amsfonts}
\usepackage{mathptmx}
\usepackage{dcolumn}
\usepackage{eucal}
\usepackage{bm}
\usepackage{color}
\usepackage[colorlinks,linkcolor=blue,citecolor=blue]{hyperref}

\usepackage{epstopdf}
\usepackage{graphicx}

\newcommand{\re}{\mathrm{Re}}
\newcommand{\im}{\mathrm{Im}}

\begin{document}
\title{ Photonic heat conduction in Josephson-coupled Bardeen-Cooper-Schrieffer superconductors}
\author{R. Bosisio}
\email{riccardo.bosisio@nano.cnr.it}
\affiliation{SPIN-CNR, Via Dodecaneso 33, 16146 Genova, Italy}
\affiliation{NEST, Instituto Nanoscienze-CNR and Scuola Normale Superiore, I-56127 Pisa, Italy}
\author{P. Solinas}
\email{paolo.solinas@spin.cnr.it}
\affiliation{SPIN-CNR, Via Dodecaneso 33, 16146 Genova, Italy}
\author{A. Braggio}
\email{alessandro.braggio@spin.cnr.it}
\affiliation{SPIN-CNR, Via Dodecaneso 33, 16146 Genova, Italy}
\author{F. Giazotto}
\email{giazotto@sns.it}
\affiliation{NEST, Instituto Nanoscienze-CNR and Scuola Normale Superiore, I-56127 Pisa, Italy}
\begin{abstract}
We investigate the photon-mediated heat flow between two Josephson-coupled Bardeen-Cooper-Schrieffer (BCS) superconductors.
We demonstrate that in standard low temperature experiments involving temperature-biased superconducting quantum interference devices (SQUIDs), this radiative contribution is negligible if compared to the direct galvanic one, but it largely exceeds the heat exchanged between electrons and the lattice phonons.
The corresponding thermal conductance is found to be several orders of magnitude smaller, for real experiments setup parameters, than the universal quantum of thermal conductance, $\kappa_0(T)=\pi k_B^2T/6\hbar$.
\end{abstract}
\maketitle
\section{Introduction}
\begin{figure}[t!]
\includegraphics[width=\columnwidth]{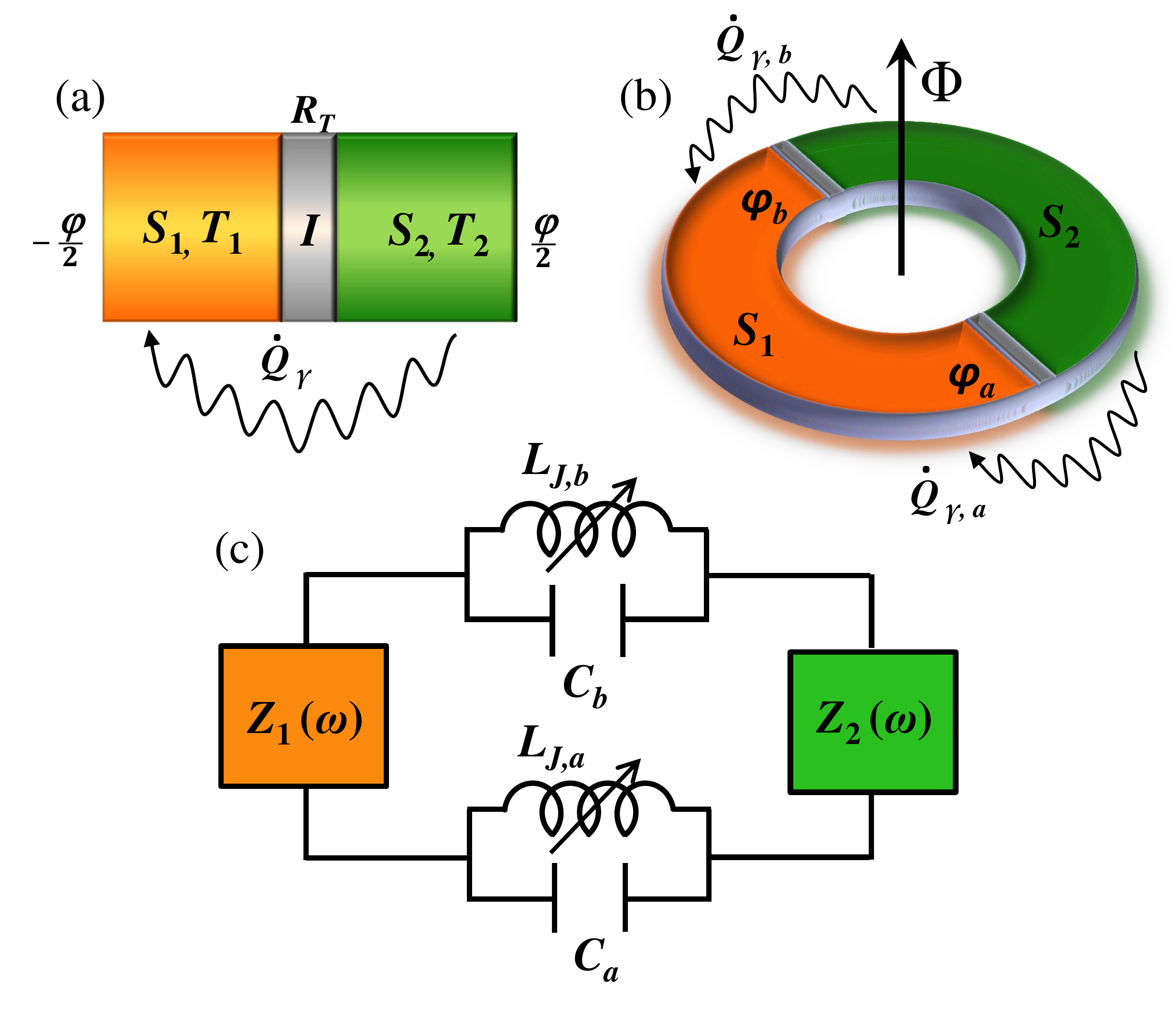}
\caption{(Color online) Sketch of the system discussed in the text and equivalent circuital description.
(a) Generic S-I-S Josephson junction, consisting of two superconductors $S_1$ and $S_2$, tunnel-coupled by an insulator \textit{I} with resistance $R_T$. $T_1$ and $T_2$ are the temperatures in $S_1$ and $S_2$, respectively, whereas $\varphi$ denotes the phase difference across the junction. The wavy line represents the photon-mediated heat exchange $\dot{Q}_\gamma$ between the two superconductors. (b) SQUID made of two Josephson junctions $a$ and $b$ and pierced by a magnetic flux $\Phi$. (c) Equivalent circuital description of the SQUID shown in (b). $S_1$ and $S_2$ are the superconducting reservoirs, and are modeled as frequency-dependent complex impedances $Z_1(\omega)$ and $Z_2(\omega)$. They are connected by two elements consisting of a Josephson inductance $L_{J,\alpha}$ and a capacitance $C_\alpha$ ($\alpha=a,b$) in parallel, describing the Josephson junctions.}
\label{fig1}
\vspace{-0mm}
\end{figure}

In the last decade the investigation of thermal transport at the mesoscale and nanoscale has known a season of very intense activity~\cite{Giazotto2006}.
In particular, many efforts have been devoted to develop solid-state electronic refrigerators using superconductors~\cite{Giazotto2006,Muhonen2012}.
In several recent experimental works~\cite{Meschke2006,Giazotto2012,MartinezPerez2014,MartinezPerez2015,Fornieri2015} the thermal properties of devices made of Josephson junctions and superconducting quantum interference devices (SQUIDs) were addressed. When a Josephson junction is thermally biased [see Fig.~\ref{fig1}(a)] it is of paramount importance to take into account all possible relaxation processes leading to heat exchange between the two superconductors $S_1$ and $S_2$. Particularly at low temperature, photon-mediated heat transport may play an important role~\cite{Meschke2006} due to the temperature dependence of its thermal conductance, $\propto T$.
This radiative contribution was experimentally investigated in the case of two normal metal resistors connected via frequency-dependent impedances~\cite{Meschke2006} represented by DC-SQUIDs and the corresponding thermal conductance between the two metal island was found to approach the universal quantum of thermal conductance $\kappa_0(T)=\pi k_B^2T/6\hbar$ at low temperatures.

In this work we focus on Josephson-coupled BCS superconductors biased in temperature as the ones used the experiments in Refs. [\onlinecite{Giazotto2012,MartinezPerez2014,MartinezPerez2015,Fornieri2015}].
Differently from Ref. [\onlinecite{Meschke2006}] where the heat exchange occurs between two normal metal, here the photonic heat exchange depends on the superconducting phase difference between the superconductors and it is a phase-coherent heat transport.
To be able to manipulate and separate directly such a contribution, we focus in particularly on the case of a temperature-biased SQUID, pierced by a magnetic flux $\Phi$.
The SQUID can be viewed as two superconducting islands (leads) characterized by frequency-dependent optical resistivities connected via frequency-dependent impedances represented by the Josephson junctions [see Fig.~\ref{fig1}(b,c)].

The paper is organized as follows. In Sec.~\ref{sec:resistivity} we review the theory of the complex resistivity $\rho(\omega)$ of a BCS superconductor; then in Sec.~\ref{sec:PhotonHeat} we derive the expression of the net heat flux $\dot{Q}_{\gamma}$ flowing via photon radiation between the superconductors of a SQUID. We discuss the associated heat conductance in Sec.~\ref{sec:K} and a realistic case in Sec.~\ref{sec:Realistic}. Finally we gather our conclusions in Sec.~\ref{sec:conclusions}.

\section{$\omega$-dependent complex resistivity}
\label{sec:resistivity}
A microwave signal affects all the charge carriers inside a superconductor.
On the one hand, the quasiparticles scatter bringing about Ohmic dissipation, and thus limiting the real part of the electric conductivity.
On the other hand, the non-dissipative, accelerative response of the Cooper pairs leads to an imaginary part of the conductivity, accounting for the superconductor kinetic inductance~\cite{Barends2009}.
This picture can be interpreted as a two-fluid model~\cite{Tinkham2012}, in which the electrodynamic response of the superconductor
results from the superposition of the condensate and the quasi-particles responses.
The microscopic treatment was developed by Mattis and Bardeen~\cite{Mattis1958}, who derived a general expression for the optical conductivity of a BCS superconductor, valid for an arbitrary degree of disorder~\cite{Mattis1958,Zimmermann1991,Pracht2013}:
\begin{equation}	
\sigma_S(\omega)=i\frac{\sigma_{dc}}{2\omega\tau}\,\times\,\left(\int_{\Delta(T)}^{\Delta(T)+\hbar\omega}I_1 \,\mathrm{d}E + \int_{\Delta(T)}^{\infty}I_2\, \mathrm{d}E \right),
\label{eq:sigma}
\end{equation}
where $\omega$ is the frequency, $\sigma_{dc}=n_ee^2\tau/m$ is the dc-conductivity of electrons with carrier density $n_e$, charge $e$, and mass $m$. The phenomenological relaxation time $\tau$ is proportional to the electron mean free path~\cite{Mattis1958,Pracht2013}, thus being a measure of the electron scattering in the superconductor. $\Delta(T)$ is the superconductor BCS energy gap at temperature $T$. The functions $I_i$ ($i=1,2$) inside the integrals are defined as:

\begin{align}
I_1=\tanh \left(\frac{E}{2k_BT}\right)&\left[(1-\aleph_{3,2})\frac{1}{P_3+P_2+i\hbar\tau^{-1}}\right.\nonumber\\
&-\left. (1+\aleph_{3,2})\frac{1}{P_3-P_2+i\hbar\tau^{-1}}\right],\\
I_2=\tanh \left(\frac{E+\hbar\omega}{2k_BT}\right)&\left[(1+\aleph_{1,2})\frac{1}{P_1-P_2+i\hbar\tau^{-1}}\right.\nonumber\\
&-\left. (1-\aleph_{1,2})\frac{1}{-P_1-P_2+i\hbar\tau^{-1}}\right]\nonumber\\
+\tanh \left(\frac{E}{2k_BT}\right)&\left[(1-\aleph_{1,2})\frac{1}{P_1+P_2+i\hbar\tau^{-1}}\right.\nonumber\\
&-\left. (1+\aleph_{1,2})\frac{1}{P_1-P_2+i\hbar\tau^{-1}}\right],
\label{eq:I1I2}
\end{align}
with
\begin{align}
&\aleph_{3,2}=\frac{\Delta(T)^2+E(E-\hbar\omega)}{P_3 P_2},\quad\aleph_{1,2}=\frac{\Delta(T)^2+E(E+\hbar\omega)}{P_1 P_2},\nonumber\\
&P_1=\sqrt{(E+\hbar\omega)^2-\Delta(T)^2},\quad P_2=\sqrt{E^2-\Delta(T)^2},\nonumber\\
&P_3=\sqrt{(E-\hbar\omega)^2-\Delta(T)^2}.\nonumber
\end{align}
In Eq.~\eqref{eq:sigma} the first integral describes the electromagnetic response of the Cooper pairs~\cite{Pracht2013}: for finite temperatures below the critical temperature $T_c$ and frequencies larger than the BCS energy gap [$\hbar\omega>2\Delta(0)$], electrons are activated across the gap by photon absorption. In particular, as it can be inferred looking at the integral boundaries, the larger $\hbar\omega$, the greater will be the range from where electrons can be excited across the gap. On the other hand, the second integral in Eq.~\eqref{eq:sigma} describes the process of radiation absorption by thermal electrons at energy $E$, which is then integrated over all energies.

Typical plots of $\re[\sigma_S]$ and $\im[\sigma_S]$ as functions of the frequency, for different temperatures $T$ and scattering time $\tau$ can be found, for instance, in Refs.~\onlinecite{Zimmermann1991,Pracht2013}.
Since we are interested in computing the overall impedance of our circuit, we rather focus on the superconductor frequency-dependent complex resistivity $\rho_S(\omega)=1/\sigma_S(\omega)$. Notice that the real and imaginary part of $\rho_S$ are related to those of $\sigma_S$ via:

\begin{subequations}\label{eq:ReImrho}
\begin{align}
&\re[\rho_S] = \frac{\re[\sigma_S]}{|\sigma_S|^2},\label{eq:ReImrho1}\\
&\im[\rho_S] = -\frac{\im[\sigma_S]}{|\sigma_S|^2}.\label{eq:ReImrho2}
\end{align}
\end{subequations}

\begin{figure}[t!]
\includegraphics[width=\columnwidth]{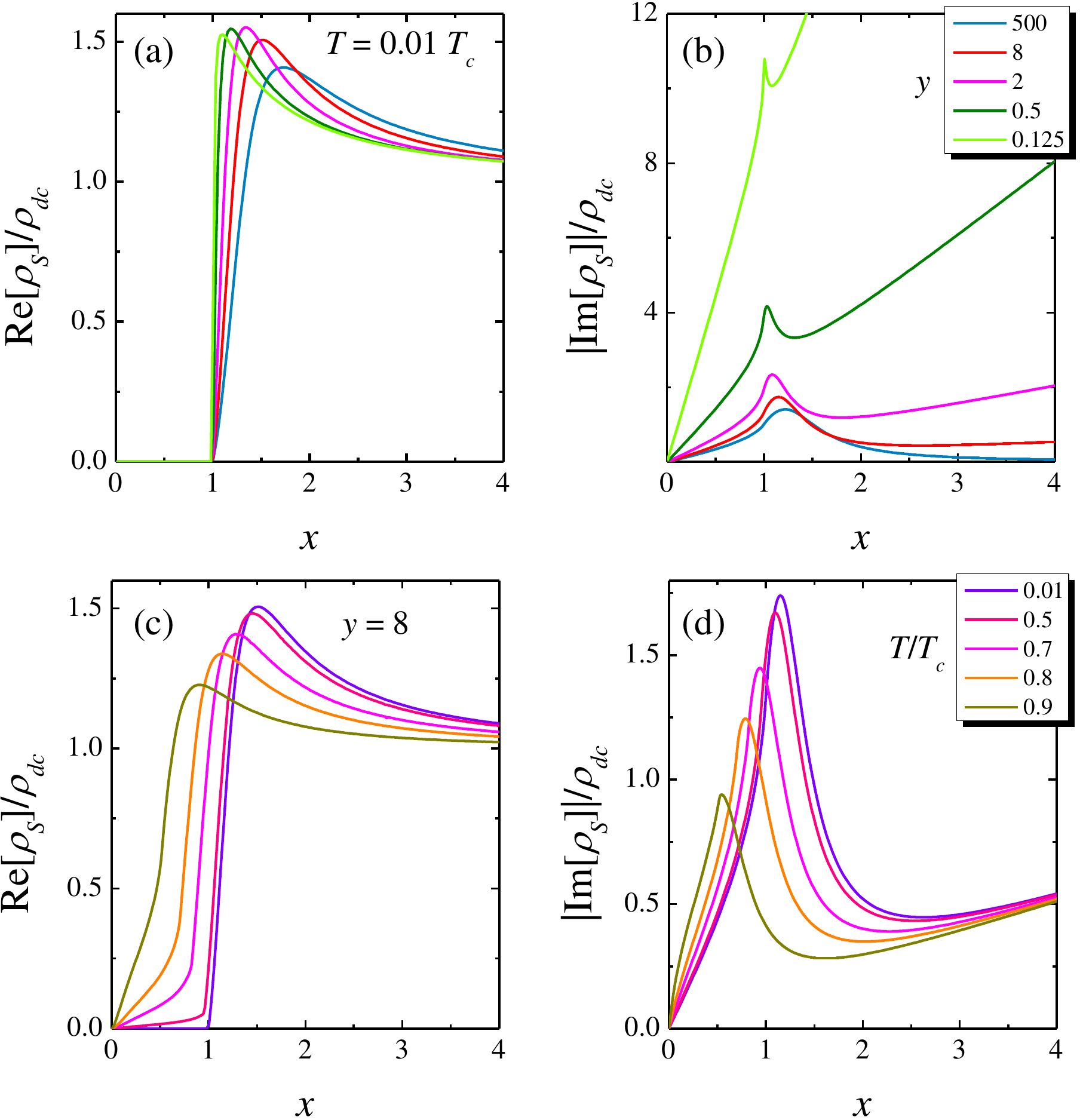}
\caption{(Color online) Complex resistivity $\rho_S(\omega)$ of a BCS superconductor, as a function of the dimensionless frequency $x=\hbar\omega/2\Delta(0)$, where $\Delta(0) \approx 1.764\,k_B T_c$ and $T_c$ is the superconducting critical temperature. Top panels: real part (a) and imaginary part (b) of $\rho_S$ at $T=0.01T_c$ for various impurity parameters $y=\hbar/2\Delta(0)\tau$ from the dirty ($y=500$) to the clean ($y=0.125$) limit. Bottom panels: real part (a) and imaginary part (b) of $\rho_S$ for $y=8$ at different temperatures. All the data are normalized by the DC resistivity $\rho_{dc}=\sigma_{dc}^{-1}$.
}
\label{fig2}
\end{figure}

Figure~\ref{fig2} shows how they behave as functions of the dimensionless frequency $x=\hbar\omega/2\Delta(0)$, for different values of the temperature $T$ and of the \textit{impurity parameter}~\cite{Zimmermann1991} $y=\hbar/2\Delta(0)\tau$.
At low temperatures ($T=0.01\,T_c$) the real (dissipative) part of $\rho_S$ is basically zero for frequencies below the gap $2\Delta(0)$ [panel (a)]. It suddenly becomes finite as $\hbar\omega>2\Delta(0)$, and its shape is only slightly altered by changing the impurity parameter from the dirty ($y=500$) to the clean ($y=0.125$) limit. By contrast, the imaginary part of $\rho_S$ is finite even at low frequencies and is much more sensitive to $y$ [panel (b)].

Panel (c) shows that $\re[\rho_S]$ is strongly dependent on the temperature for a fixed intermediate disorder $y=8$. At low frequency ($x<1$), increasing the temperature makes $\re[\rho_S]$ to become finite, whereas at higher frequency its effect is smoothed. 
Finally, in panel (d) we see that by increasing $T$ the maximum value of $\im[\rho_S]$ decreases; on the other hand, we observe that in the limit $x\ll 1$ its slope increases with $T$, in line with the behavior of the the superconductor kinetic inductance (see related discussion in the next section).

\section{Photon-mediated heat exchange}
\label{sec:PhotonHeat}
In order to calculate the photon-mediated heat exchange, we need to compute the impedances of all the elements of the circuit [see Fig.~\ref{fig1}(c)].
The two BCS superconductors of the SQUID have frequency-dependent complex impedances which are proportional to their resistivity via the Ohm's law: 
\begin{equation}
Z_i(\omega)=\rho_{S,i}(\omega)\,\frac{\ell_i}{\mathcal{A}_i},\qquad (i=1,2)
\label{eq:Rw}
\end{equation}
$\ell_i$ being the superconductor $S_i$ length and $\mathcal{A}_i$ its cross-sectional surface perpendicular to the current flow.
The relation above is valid if the relevant frequency scales are such that the radiation wavelength $\lambda$ is larger than the system size.
The relevant frequencies for heat exchange between the two superconductors are those around the gap ($x\sim 1$, see Fig.~\ref{fig2}).
For Aluminum, a frequency $\hbar\omega$ of the order of the BCS gap $2\Delta(0)\approx 362~\mu$eV corresponds to a wavelength $\lambda\simeq 3.4\times 10^{-3}~$m, which is much larger than the typical system size $\ell\sim 10^{-5}~$m (see Sec.~\ref{sec:K}). The response of the system is thus local even at larger frequencies, and Eq.~\eqref{eq:Rw} holds.

The real part of $Z_i(\omega)$ (the resistance) is dissipative;
its imaginary part (the reactance) is non-dissipative: in linear response (for small frequencies $\omega$), it can be related to the superconductor kinetic inductance $L_{k,i}$ via:
\begin{equation}
\im[Z_i(\omega)]\simeq\omega\,L_{k,i},\qquad (i=1,2).
\label{eq:Lkin}
\end{equation}
%

According to the circuital description of Fig.~\ref{fig1}(c), $S_1$ and $S_2$ are connected via frequency-dependent impedances $Z_a(\omega)$ and $Z_b(\omega)$ associated to the Josephson junctions $a$ and $b$. 
Whereas in the most general situation one could compute the \textit{full} impedance of the junctions~\cite{Catelani2011,Pop2014}, there are strong experimental and theoretical indications~\cite{Meschke2006,Ojanen2008,Pascal2011,Golubev2013} that their behavior can be very well described by purely reactive \textit{LC} elements, made of a capacitor $C_\alpha$ in parallel with a Josephson inductance
\begin{equation}
L_{J,\alpha}=\frac{\Phi_0}{2\pi}\left(\frac{\partial I_{J,\alpha}}{\partial\varphi_\alpha}\right)^{-1} = \frac{\Phi_0}{2\pi} \frac{1}{I_{c,\alpha}\cos\varphi_\alpha},
\label{eq:LJ}
\end{equation}
with $\alpha=a,b$, $\Phi_0$ being the flux quantum, and $\varphi_\alpha$ the junction phase difference. In the following we will use this simplified but realistic model. The Josephson current $I_{J,\alpha}$ through the junction is assumed to be sinusoidal~\cite{Tinkham2012}, $I_{J,\alpha}=I_{c,\alpha} \sin\varphi_\alpha$, with $I_{c,\alpha}$ the maximum Josephson current given by~\cite{Giazotto2005,Tirelli2008}:
\begin{align}
I_{c,\alpha}=&\frac{1}{2eR_{T,\alpha}}\Big|\int_{-\infty}^\infty d\varepsilon \{[1-2f(\varepsilon,T_1)]\re[\mathcal{F}_{S_1}(\varepsilon)]\im[\mathcal{F}_{S_2}(\varepsilon)]\cr
&+[1-2f(\varepsilon,T_2)]\re[\mathcal{F}_{S_2}(\varepsilon)]\im[\mathcal{F}_{S_1}(\varepsilon)]\}\Big|.
\label{eq:Icalpha}
\end{align}
Here $R_{T,\alpha}$ is the junction $\alpha$ tunnel resistance, $f(\varepsilon,T)$ is the Fermi-Dirac distribution at temperature $T$, and $\mathcal{F}_{S_i}(\varepsilon) = \Delta(T_i)/\sqrt{(\varepsilon+i\Gamma_i)^2-\Delta^2(T_i)}$, $\Gamma_i$ being the Dynes\footnote{Equation~\eqref{eq:Icalpha} is valid in general even if the superconductors $S_1$ and $S_2$ have different gaps, $\Delta_1\neq \Delta_2$. Since in this work we do not consider such a situation, in the following we will refer to a unique function $\Delta(T)$ with no ambiguity.} parameter~\cite{Dynes1978}.
Since the superconducting gap $\Delta$ is temperature-dependent, we see that $I_{c,\alpha}$ is a function of both $T_1$ and $T_2$. If the latter are equal, $T_1=T_2$, the above expression reduces to the well-known Ambegaokar-Baratoff formula~\cite{Tinkham2012}:
\begin{equation}
I_{c,\alpha}=\frac{\pi\Delta(T)}{2eR_{T,\alpha}} \tanh\left[\frac{\Delta(T)}{2k_BT}\right].
\label{eq:Icalpha2}
\end{equation}
The Josephson junction impedance $Z_\alpha(\omega)$ is found by:
\begin{equation}
 \frac{1}{Z_\alpha(\omega)}= \,i\omega C_\alpha+\frac{1}{i\omega L_{J,\alpha}},
\end{equation}
and using Eq.~\eqref{eq:LJ} we obtain 
\begin{equation}
Z_\alpha(\omega) = \frac{-i\omega}{\omega^2C_\alpha-(2\pi/\Phi_0) I_{c,\alpha}\cos\varphi_\alpha}.
\label{eq:Zalpha}
\end{equation}


%
Since the Josephson inductances $L_{J,\alpha}$ typically dominate the superconductors kinetic inductances $L_{k,i}$, the superconducting phase drops $\varphi_\alpha$ take place entirely across the corresponding junctions.
These phase drops $\varphi_a$ and $\varphi_b$ are then related via the fluxoid quantization~\cite{Tinkham2012} $\varphi_a+\varphi_b+2\pi\Phi/\Phi_0=2k\pi$ (with $k$ integer), where $\Phi$ is the external magnetic flux piercing the SQUID loop [see Fig.~\ref{fig1}(b)].
Moreover, for any given magnetic flux $\Phi$ and temperature $T$, the conservation of the circulating supercurrent along the loop reads $I_{c,a}\sin\varphi_a=I_{c,b}\sin\varphi_b$, with $I_{c,\alpha}$ given by Eq.~\eqref{eq:Icalpha}.
These two expressions allow us to rewrite the phases $\varphi_a$ and $\varphi_b$ in terms of the external magnetic flux $\Phi$ and a dimensionless ratio between the junctions critical currents, $r=I_{c,a}/I_{c,b}= R_{T,b}/R_{T,a}$ via the following mapping:
\begin{subequations}\label{eq:cosines}
\begin{align}
& \cos\varphi_a=\frac{r+\cos(2\pi\Phi/\Phi_0)}{\sqrt{1+r^2+2r\cos(2\pi\Phi/\Phi_0)}},\label{eq:cosinesa}\\
& \cos\varphi_b=\frac{1+r\cos(2\pi\Phi/\Phi_0)}{\sqrt{1+r^2+2r\cos(2\pi\Phi/\Phi_0)}}.\label{eq:cosinesb}
\end{align}
\end{subequations}
Substituting back into Eqs.~\eqref{eq:Zalpha} gives:
%
%
%
\begin{subequations}\label{eq:Z}
\begin{align}
& Z_a(\omega)=\frac{i\omega\sqrt{1+r^2+2r\cos\phi}}{(2\pi/\Phi_0) I_{c,a}\left(r+\cos\phi\right)-\omega^2C_a\sqrt{1+r^2+2r\cos\phi}},\label{eq:Za}\\
& Z_b(\omega)=\frac{i\omega\sqrt{1+r^2+2r\cos\phi}}{(2\pi/\Phi_0) I_{c,b}\left(1+r\cos\phi\right)-\omega^2C_b\sqrt{1+r^2+2r\cos\phi}},\label{eq:Zb}
\end{align}
\end{subequations}
where $\phi=2\pi\Phi/\Phi_0$.
The total series impedance of the circuit is found by summing all the impedances of the chain,
\begin{equation}
Z_{\mathrm{tot}}(\omega) =  Z_a(\omega)+Z_1(\omega)+Z_b(\omega) +Z_2(\omega).
\label{eq:Ztot}
\end{equation}
It is interesting to emphasize the behavior of the junction impedances $Z_\alpha(\omega)$ ($\alpha=a,b$) in the particular situation of a symmetric SQUID ($r=1$) when the magnetic flux is half the flux quantum, $\Phi=\Phi_0/2$, and hence $\cos\phi=-1$. At this singular point the Josephson inductance $L_{J,\alpha}$ diverges, and Eqs.~\eqref{eq:Z} reduce to
\begin{equation}
Z_\alpha(\omega) =  \frac{1}{i\omega C_\alpha}\quad\text{with}\quad \alpha=a,b,
\label{eq:Zshunt}
\end{equation}
that is, the junction junctions are completely shunted by $C_\alpha$, and their impedances become purely \textit{capacitive}.

We now review how to calculate the photon-mediated heat flux between the two superconductors $S_1$ and $S_2$. According to the fluctuation-dissipation theorem~\cite{Cleland2004,Callen1951,Lifshitz1998}, this quantity is the difference between the electromagnetic noise power radiated by them.
To calculate it, we start by recalling that the voltage noise density produced by $S_i$ at temperature $T_i$ is proportional to the dissipative component of its impedance via~\cite{Cleland2004}
\begin{equation}
\mathcal{S}_{V,i}(\omega)= 4\hbar\omega \re[Z_i(\omega)] \left[n(\hbar\omega,T_i)+\frac{1}{2}\right],
\label{eq:SV}
\end{equation}
where $n(\epsilon,T)=[\mathrm{exp}(\epsilon/k_B T)-1]^{-1}$ is the Bose-Einstein distribution of photons at temperature $T$, while the factor $1/2$ is the zero-point term, that will cancel out in writing the \textit{net} transferred heat between $S_1$ and $S_2$.
The current noise through the circuit due to the voltage produced by $Z_i$ is then~\cite{Cleland2004}
\begin{equation}
\mathcal{S}_{I,i}(\omega)=\frac{\mathcal{S}_{V,i}(\omega)}{\left|Z_\mathrm{tot}(\omega)\right|^2}.
\label{eq:SI}
\end{equation}
Now, the power density produced by the noise of $Z_1$ and dissipated into $Z_2$ is 
\begin{equation}
\mathcal{S}_{P,12}(\omega)=\re[Z_2(\omega)]\mathcal{S}_{I,1}(\omega), 
\label{eq:SP}
\end{equation}
so that the corresponding dissipated power is found by integrating over all the frequencies.
%
%
The net heat flux flowing by photon radiation into $S_1$ from the superconductor $S_2$ is finally:
\begin{equation}
\dot{Q}_{\gamma}= \int_0^\infty\frac{\mathrm{d}\omega}{2\pi}\,\hbar\omega\,\mathcal{T}(\omega,T_1,T_2)\left[n(\hbar\omega,T_2)-n(\hbar\omega,T_1)\right],
\label{eq:Pnet}
\end{equation}
where we have defined the effective photon transmission coefficient
\begin{equation}\label{eq:transm}
\mathcal{T}(\omega,T_1,T_2)=\frac{4\re[Z_1(\omega,T_1,T_2)]\re[Z_2(\omega,T_1,T_2)]}{\left|Z_\mathrm{tot}(\omega,T_1,T_2)\right|^2}.
\end{equation}
Notice also that we have assumed the convention by which we consider as positive the heat injected from $S_2$ into $S_1$ when $T_2>T_1$.
%
%
\section{Photonic Heat Conductance}
\label{sec:K}
%
%
It is instructive to start our discussion by analyzing the heat transport within the linear response approximation.
When the difference between the superconductors temperatures is sufficiently small, we can write
\begin{equation}
\dot{Q}_{\gamma} = \kappa_\gamma(T)\,\delta T,
\label{eq:Pnet_lin}
\end{equation}
where we have defined $\kappa_\gamma(T)$ as the thermal conductance associated to the photon-mediated heat transport, and $\delta T\equiv T_2-T_1\ll T\equiv (T_1+T_2)/2$.
Expanding the difference between the Bose distributions in Eq.~\eqref{eq:Pnet} to the first order in $\delta T$, we can write an expression for $\kappa_\gamma$:
\begin{equation}
\kappa_\gamma(T)=\int_0^\infty\,\frac{\mathrm{d}\omega}{2\pi}\frac{(\hbar\omega)^2}{k_B T^2}\, \mathcal{T}(\omega,T)\,\sinh^{-2}\left(\frac{\hbar\omega}{2k_B T}\right).
\label{eq:kgamma}
\end{equation}
In the following we assume the temperatures of the two superconductors $S_1$ and $S_2$ to be equal, $T_1=T_2\equiv T$ and investigate the heat conductance normalized by the universal quantum $\kappa_0(T)=\pi k_B^2T/6\hbar$:
\begin{equation}
\mathcal{K}_\gamma(T)=\frac{\kappa_\gamma(T)}{\kappa_0(T)}.
\label{eq:Krenorm}
\end{equation}
All the results discussed below and in the next section are calculated for a SQUID made of Al/AlOx/Al Josephson junctions: we assume the superconductors typical length $\ell_i$ and cross-section area $\mathcal{A}_i$ to be $10~\mu$m and $200~$nm$\times 20~$nm, respectively, the impurity parameter $y=8$, the Aluminum critical temperature $T_c=1.19$~K, and dc-conductivity~\cite{Ashcroft1976} $\sigma_{dc}=2.8\times10^7~$S/m. Furthermore, we assume $S_1$ and $S_2$ to have the same zero-temperature BCS gap $\Delta(0) \approx 1.764\,k_B T_c\simeq 181~\mu$eV~\cite{Tinkham2012}.
%
%
\begin{figure}[t!]
\includegraphics[width=\columnwidth]{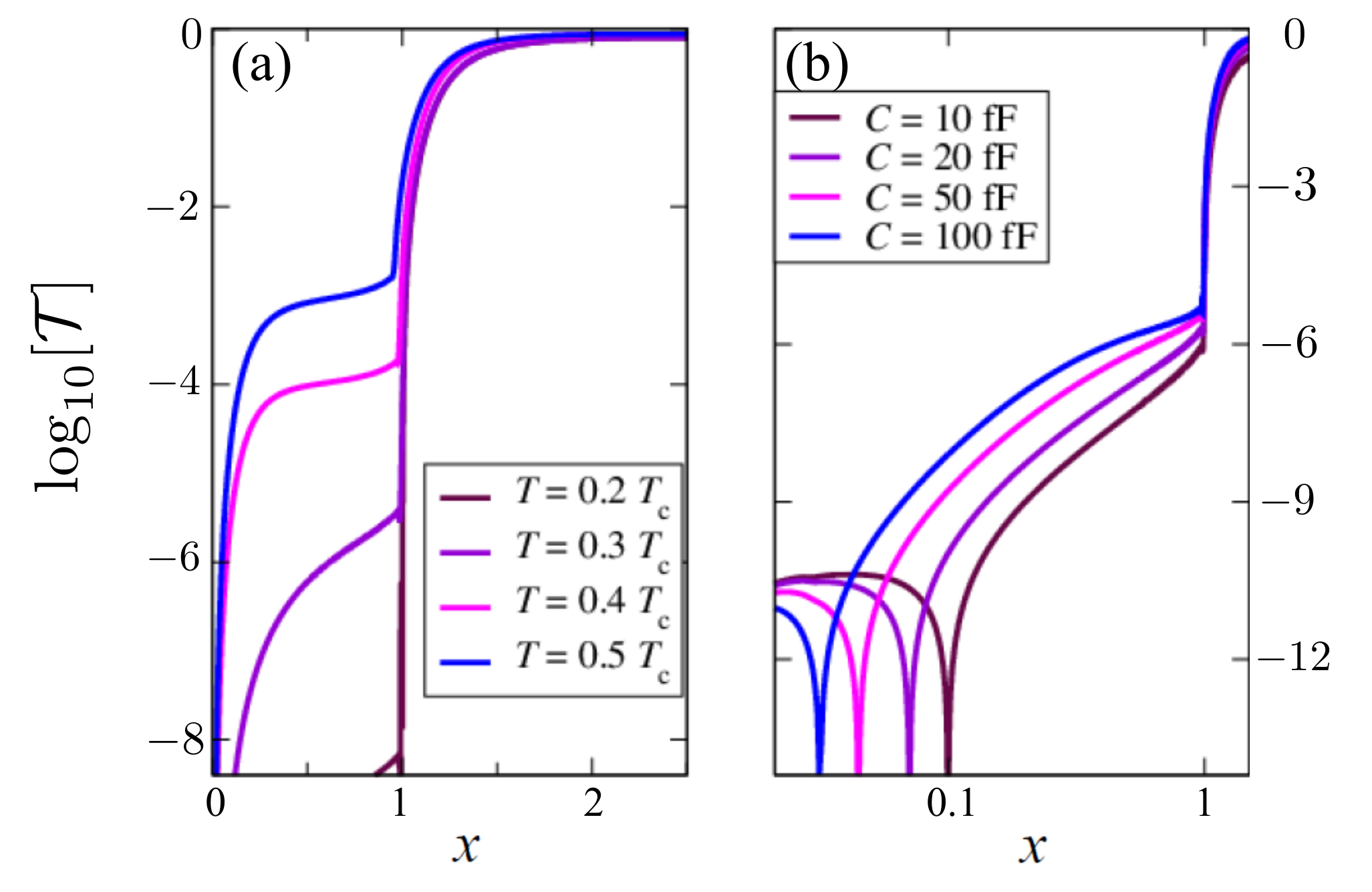}
\caption{(Color online) Effective photon transmission coefficient $\mathcal{T}$ from Eq.~\eqref{eq:transm} as function of the dimensionless frequency $x=\hbar\omega/2\Delta(0)$ for a symmetric SQUID ($r=1$) with $T_1=T_2\equiv T$. (a) Curves for different temperatures, and for equal junction capacitances $C_a=C_b=50~$fF. (b) Curves for different junctions capacitances, and for a temperature $T=0.3 T_c$. In both panels the critical currents are $I_{c,a}(0)=I_{c,b}(0)=10~$nA and the magnetic flux is $\Phi=0$.}
\label{fig3}
\end{figure}
%
%

Figure~\ref{fig3} shows the effective photon transmission coefficient $\mathcal{T}(\omega,T)$ appearing in Eq.~\eqref{eq:kgamma}, obtained from the expression~\eqref{eq:transm} when $T_1=T_2\equiv T$ for a symmetric SQUID ($r=1$). In panel (a) we see that $\mathcal{T}$ strongly depends on the temperature at frequencies below the gap ($x<1$), being essentially zero for sufficiently low $T$ (see for instance the curve for $T=0.2T_c$), whilst progressively increasing for larger $T$. 
Moreover, at large frequency ($x\gg 1$) the effect of the temperature on the BCS complex resistivity become negligible [see Fig.~\ref{fig2}(c,d)], and this translates directly on the behavior of $\mathcal{T}$: all the curves tend to collapse on a single one approaching unitary transmission.
Panel (b) shows the effect of the junctions capacitance on $\mathcal{T}$. Let us consider the case when $\Phi=0$: equations~\eqref{eq:Z} can be written in the form
\begin{equation}
Z_\alpha=-i\,\frac{\omega/C_\alpha}{\omega^2-\omega_0^2}\quad\text{with}\quad \omega_0\equiv \frac{(2\pi/\Phi_0)I_{c,\alpha}}{C_\alpha}.
\label{eq:Zresonance}
\end{equation}
From these expressions we see that at resonance ($\omega=\omega_0$) the impedance diverges, and $\mathcal{T}(\omega,T)$ vanishes.
By increasing $C_\alpha$, this resonance frequency $\omega_0$ is red-shifted.
Moreover, for $\omega_0>\omega$, the absolute value of $Z_\alpha$ is a monotonically decreasing function of $C_\alpha$. 
Since in Eq.~\eqref{eq:Ztot} the superconductor impedances $Z_1$ and $Z_2$ do \textit{not} depend on the junction capacitance $C_\alpha$, we conclude that increasing the latter lowers the total impedance $Z_\text{tot}$, thereby resulting in a larger photon transmission $\mathcal{T}$ [Eq.~\eqref{eq:transm}].
All these considerations are confirmed by looking at the different curves in Fig.~\ref{fig3}(b).

\begin{figure}[t!]
\includegraphics[width=\columnwidth]{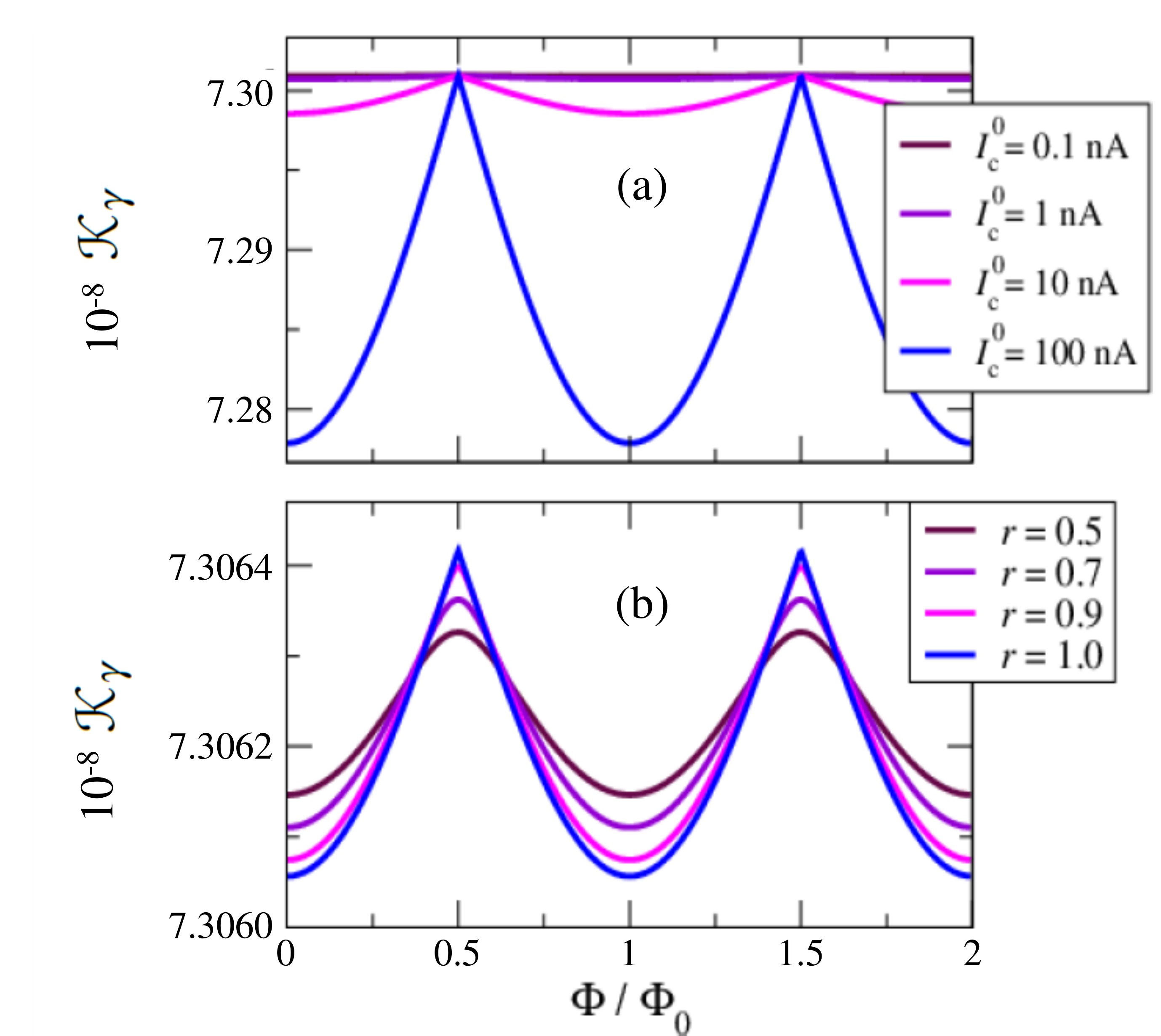}
\caption{(Color online) Normalized heat conductance $\mathcal{K}_\gamma$ from Eq.~\eqref{eq:Krenorm} as function of the external magnetic flux $\Phi$. (a) Symmetric SQUID ($r=1$) and different values of the junctions critical current $I_{c,a}(0)=I_{c,b}(0)\equiv I_c^0$. (b) Asymmetric SQUID ($r\neq 1$) for a given value of $I_{c,b}^0=1$~nA. Other parameters are the temperature $T_1=T_2\equiv T=0.3T_c$, and the junctions capacitance $C_a=C_b\equiv C=50$~fF.}
\label{fig4}
\end{figure}

Figure~\ref{fig4} shows $\mathcal{K}_\gamma$ as function of the external magnetic flux $\Phi$.
We first observe that $\mathcal{K}_\gamma$ is a very small quantity, that is, the photonic heat conductance $\kappa_\gamma(T)$ is typically several orders of magnitude smaller than the quantum of heat conductance, $\kappa_0(T)$.
Next we see that $\mathcal{K}_\gamma$ is periodic with respect to $\Phi$, as it could be inferred from Eqs.~\eqref{eq:Z} to~\eqref{eq:Pnet}.
In panel (a) we consider a symmetric SQUID ($r=1$) for different values of the zero-temperature junctions critical current $ I_c^0\equiv I_{c,a}(0)=I_{c,b}(0)$.
Increasing $I_c^0$ enhances the Josephson coupling between $S_1$ and $S_2$, and thus results in a larger modulation amplitude of the heat conductance. The lowest values (0.1~nA and 1~nA) curves appear basically indistinguishable and their modulation amplitude is very small, if compared to the $I_c^0=100$~nA case.
By contrast, we notice that the maximum value of $\mathcal{K}_\gamma$, attained for multiples of $\Phi_0/2$, remains constant irrespective of $I_c^0$. This can be understood by looking at Eq.~\eqref{eq:Zshunt}: when $\Phi=\Phi_0/2$ the junctions are capacitively shunted, the total impedance $Z_\text{tot}$ [Eq.~\eqref{eq:Ztot}] becomes completely independent of the critical current, and so does the effective photon transmission coefficient $\mathcal{T}$ [Eq.~\eqref{eq:transm}]. The same behavior is thus reflected on the heat conductance.
%
%

In Fig.~\ref{fig4}(b) we investigate the effect of an asymmetry between the two Josephson junctions in the SQUID. More specifically, they may have different tunnel resistances (and hence different critical currents): this is quantified by the parameter $r$ introduced in Eqs.~\eqref{eq:cosines}. Assuming a certain value for one of the junctions critical currents at zero temperature, say $I_{c,b}^0=1$~nA, we see that the more asymmetric the SQUID, the smaller the modulation contrast of $\mathcal{K}_\gamma$ will be:
the symmetric case ($r=1$) is thus the one with the highest modulation visibility, corresponding to the best impedance-matched condition.
We notice also that the cusps which can be observed in the symmetric case at multiples of $\Phi_0/2$ are smoothed by the asymmetry: the $r=0.5$ curve, for instance, exhibits a more regular sinusoidal shape.

\begin{figure}[t!]
\includegraphics[width=\columnwidth]{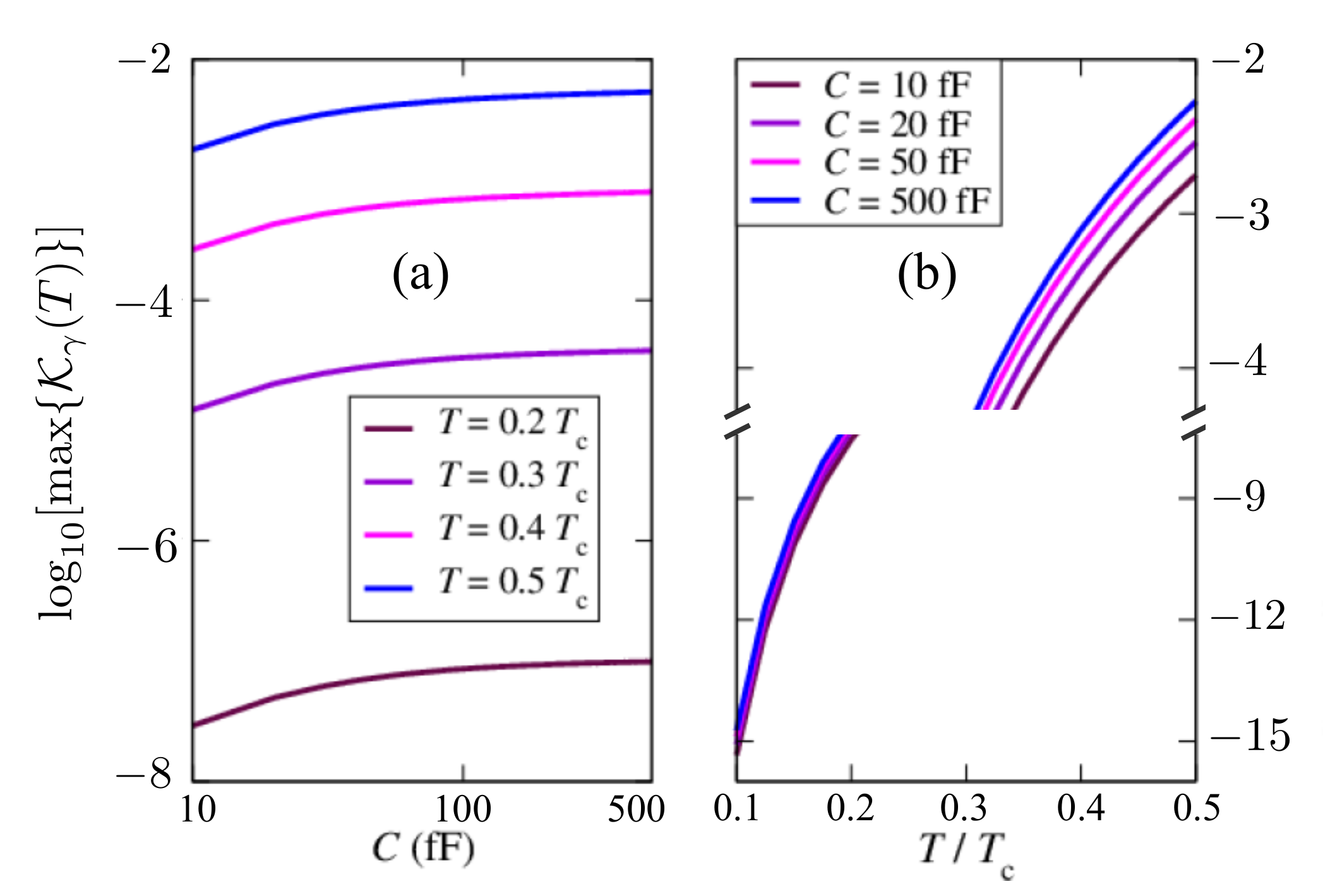}
\caption{(Color online) Maximum value of the normalized heat conductance $\max\{\mathcal{K}_\gamma \}$ as a function of the junctions capacitance $C\equiv C_a=C_b$ (a) and of the superconducting leads temperature $T\equiv T_1=T_2$ (b). The calculations have been made for a symmetric SQUID ($r=1$) assuming a junction critical current $I_{c}^0=10$~nA, and having set the external magnetic flux to $\Phi_\mathrm{ext}=\Phi_0/2$.}
\label{fig5}
\end{figure}

Figure~\ref{fig5} shows how the \emph{maximum} value of the normalized heat conductance, $\max\{\mathcal{K}_\gamma\}$, obtained when $\Phi_\mathrm{ext}=\Phi_0/2$ (see Fig.~\ref{fig4}), varies with the junctions capacitance $C_a=C_b\equiv C$ and with the superconducting leads temperature $T_1=T_2\equiv T$. From panel (a) we see that raising $C$ from 10~fF to 500~fF leads to an enhancement of $\max\{\mathcal{K}_\gamma\}$ of about one order of magnitude in the range of temperatures considered.
This is a direct consequence of the behavior of the effective photon transmission coefficient investigated in Fig.~\ref{fig3}.
There we saw that, if $\omega>\omega_0$, $\mathcal{T}$ it is a monotonically increasing function of $C$, and so is $\mathcal{K}_\gamma$, by virtue of Eq.~\eqref{eq:kgamma}.
A larger junction capacitance leads to a stronger coupling between $S_1$ and $S_2$, which in turn leads to a larger heat exchanged between them.

The effect of the temperature is much more pronounced: panel (b) shows that raising the temperature from $0.1T_c$ to $0.5T_c$ leads to a drastic increase of $\max\{\mathcal{K}_\gamma \}$ of up to thirteen orders of magnitude! This exponential increase can be inferred by looking at Eq.~\eqref{eq:kgamma}. The transmission coefficient $\mathcal{T}$ is multiplied by a weight function $[(\hbar\omega)^2/k_BT^2]\,\sinh^{-2}(\hbar\omega/2k_BT)$. The latter is a bell-shaped function of $\omega$, centered at $\omega=0$, and whose width is proportional to $T$. The effect of this weight function is thus to select the frequency window in which the contribution of $\mathcal{T}$ is most relevant: for sufficiently large $T$, it encompasses the exponential increase of $\mathcal{T}$ which takes place for $\omega>\omega_0$ [see Fig.~\ref{fig3}(b)]. The larger $T$, the stronger the exponential increase ``seen'' by the weight function will be.

On these scales, changing the junctions capacitance does not lead to appreciable effects: the various curves in panel (b) can be well distinguished only at large $T$, whereas they tend to collapse on a single curve as $T$ is lowered.
We then conclude that the photon-mediated contribution to the heat conduction in a Josephson device is extremely sensitive to the temperature, and becomes more important when the latter is high.
These behaviors for the heat conductance are the same for all the range of parameters considered, and $\kappa_\gamma$ always remains several orders of magnitude smaller than the quantum of heat conductance.
%
%
\section{Net photonic heat transfer in a realistic case}
\label{sec:Realistic}
We now consider a SQUID made of two superconductors held at different temperatures, and we use Eq.~\eqref{eq:Pnet} to evaluate the \emph{net} heat radiated between them when one of the two (say, $S_1$) is cooled down ($T_1<T_2$).

\begin{figure}[t!]
\begin{center}
\includegraphics[width=\columnwidth]{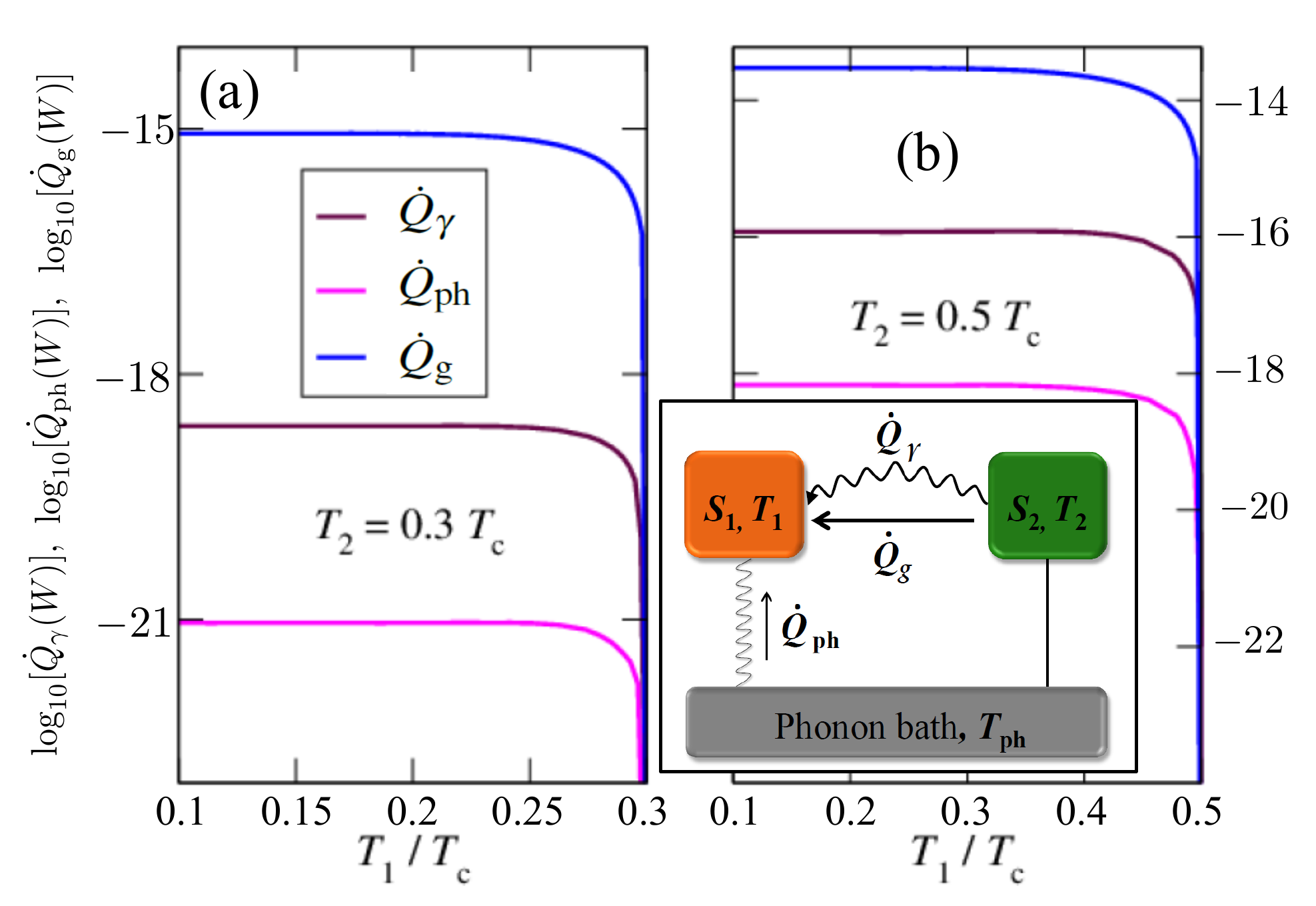}
\caption{(Color online) Net photon-mediated heat $\dot{Q}_{\gamma}$ transferred between $S_1$ and $S_2$, phonon-mediated heat $\dot{Q}_\text{ph}$ exchanged between $S_1$ and the lattice, and direct galvanic heat $\dot{Q}_\text{g}$, as functions of $T_1$, for $T_2=0.3T_c$ (a) and $T_2=0.5T_c$ (b). Inset: schematic of the thermal model. The superconducting lead $S_1$ exchanges heat by electromagnetic radiation and by direct galvanic coupling with $S_2$, and by electron-phonon interaction with its lattice phonons. Parameters: $I_{c,b}^0=10$~nA, $I_{c,a}^0=r I_{c,b}^0$, $r=0.95$, $\Phi=\Phi_0/2$, $C_a=C_b=C=50$~fF, $\Gamma_1=10^{-4}\Delta(0)$, $\Sigma=2\times 10^8~$W$\,$K$^{-5}\,$m$^{-3}$, and $\mathcal{V}=4\times10^{-19}~$m$^{3}$.}
\label{fig6}
\end{center}
\end{figure}

In a realistic experimental situation, there will obviously be a galvanic heat flow $\dot{Q}_\text{g}$ from the hot superconductor into the cold one. Besides, the latter will also exchange heat $\dot{Q}_\text{ph}$ with its lattice phonons~\cite{Timofeev2009ph,Maisi2013} (see inset of Fig.~\ref{fig6}). 
The galvanic heat flow contains two terms, a quasi-particle one, and a second one due to the interference between quasiparticles and Cooper pairs condensate flowing through each junction of the SQUID~\cite{Giazotto2012apl}:
\begin{equation}
\dot{Q}_\text{g}=\dot{Q}_\text{qp}(T_1,T_2)-\dot{Q}_\text{int}(T_1,T_2,\Phi,r).
\label{eq:Qg}
\end{equation}
Here both $\dot{Q}_\text{qp}$ are sums of each junction contributions, $\dot{Q}_\text{qp}=\dot{Q}_\text{qp}^a+\dot{Q}_\text{qp}^b$, and $\dot{Q}_\text{int}=\dot{Q}_\text{int}^a \cos\varphi_a+\dot{Q}_\text{int}^b\cos\varphi_b$. Using Eqs.~\eqref{eq:cosines}, the latter can be written as
\begin{equation}
\dot{Q}_\text{int}(T_1,T_2,\Phi,r)=\dot{Q}_\text{int}^b\sqrt{1+r^2+2r\cos\left(2\pi\frac{\Phi}{\Phi_0}\right)}.
\end{equation}
In the above expressions $\dot{Q}_\text{qp}^\alpha$ and $\dot{Q}_\text{qp}^\alpha$ are given by~\cite{Maki1965,Guttman1997,Guttman1998,Zhao2003,Zhao2004}:
\begin{align}
&\dot{Q}_\text{qp}^\alpha=\frac{1}{R_{T,\alpha}}\int_0^{\infty}dE E \mathcal{N}_1(E)\mathcal{N}_2(E)[f(E,T_1)-f(E,T_2)],\nonumber\\
&\dot{Q}_\text{int}^\alpha=\frac{1}{R_{T,\alpha}}\int_0^{\infty}dE E \mathcal{M}_1(E)\mathcal{M}_2(E)[f(E,T_1)-f(E,T_2)],\nonumber
\end{align}
where $\mathcal{N}_i(E)=\left|\re\left[(E+i\Gamma_i)/\sqrt{(E+i\Gamma_i)^2-\Delta^2(T_i)}\right]\right|$ is the smeared BCS density of states of superconductor $S_i$ at temperature $T_i$, $\mathcal{M}_i(E)=\left|\im\left[\Delta(T_i)/\sqrt{-(E+i\Gamma_i)^2+\Delta^2(T_i)}\right]\right|$, and $\alpha=a,b$.

The phonon-mediated contribution in a superconductor is given by~\cite{Maisi2013}:
\begin{align}
\dot{Q}_\text{ph}=&\frac{\Sigma \mathcal{V}}{24\zeta(5)k_B^5}\int_0^\infty d\varepsilon \epsilon^3[n(\epsilon,T_1)-n(\epsilon,T_\text{ph})]\cr
& \times \int_{-\infty}^\infty dE \mathcal{N}_1(E) \mathcal{N}_1(E+\epsilon)\left[1-\frac{\Delta^2(T_1)}{E(E+\epsilon)}\right]\cr
& \times [f(E,T_1)-f(E+\epsilon,T_1)],
\label{eq:Qph}
\end{align}
where $\Sigma$ is the material constant for electron-phonon coupling~\cite{Giazotto2006}, $\mathcal{V}$ is the volume of the superconducting electrodes, and $\zeta(x)$ is the Riemann zeta function. Phonons in $S_1$ are assumed to be in equilibrium at temperature $T_\text{ph}$, with occupation given by the Bose-Einstein distribution $n(\epsilon,T_\text{ph})$. Furthermore, we assume that lattice phonons present in every part of the SQUID are fully thermalized with the substrate phonons residing at $T_\text{ph}$, thanks to the vanishingly small Kapitza resistance between thin metallic films and the substrate at low temperatures~\cite{Giazotto2012,Wellstood1994}.

Figure~\ref{fig6} shows the comparison between the net photon-mediated heat $\dot{Q}_{\gamma}$ flowing into $S_1$ from $S_2$ [Eq.~\eqref{eq:Pnet}], the heat $\dot{Q}_\text{ph}$ exchanged between $S_1$ and the lattice phonons [Eq.~\eqref{eq:Qph}], and the galvanic heat flow $\dot{Q}_\text{g}$ [Eq.~\eqref{eq:Qg}]. The system considered is a SQUID as in Fig.~\ref{fig1}(b) with a small asymmetry $r=0.95$ between the two Josephson junctions.
The phonons and quasiparticles in $S_2$ are assumed to be at the same temperature, $T_{\rm{ph}}=T_2$, while the temperature of $S_1$ is gradually decreased from $T_1=T_2=0.3 T_c$ [panel (a)] or from $T_1=T_2=0.5 T_c$ [panel (b)] down to $T_1=0.1 T_c$.
Looking at the figure from right to left, all the three terms are initially zero at $T_1=T_2$, then increase as $T_1$ is lowered, and finally tend to saturate for sufficiently low $T_1$.
We notice that $\dot{Q}_{\gamma}$ dominates $\dot{Q}_\text{ph}$ in the whole range of temperature considered, and is in turn dominated by $\dot{Q}_\text{g}$. 
This is interesting because it shows that, if the photon-mediated heat flow is a small correction ($\lesssim 1\%$) with respect to the leading galvanic contribution, it is however much more important than the phonon-mediated one.
By contrast, we recall that in normal metals the heat exchanged between electrons and phonons is typically much larger~\cite{Giazotto2006}, and thus for these systems the above statements may not hold.
We point out that the data shown in Fig.~\ref{fig6} have been calculated at a fixed magnetic flux $\Phi=\Phi_0/2$ that maximizes $\dot{Q}_{\gamma}$. Changing $\Phi$ also impacts $\dot{Q}_\text{g}$ through the interference term $\dot{Q}_\text{int}$ [see Eq.~\eqref{eq:Qg}]: we have done extensive numerical simulations for different values of $\Phi$ and checked that the results are only slightly altered, the qualitative picture being essentially unchanged.
\\

We stress that the amount of heat radiated between the two superconductors in a SQUID setup as the one considered in this work [Fig.~\ref{fig1}(b)] is much larger than it would be in the case of a single Josephson junction setup [Fig.~\ref{fig1}(a)]. The latter situation would indeed correspond to a mismatched geometry~\cite{Timofeev2009} in which $S_1$ and $S_2$, connected by a single \textit{LC} parallel as the ones in Fig.~\ref{fig1}(c), are then grounded through some small stray capacitances $C_g$. The total impedance of the circuit would include the contributions associated to $C_g$ and would be much larger than Eq.~\eqref{eq:Ztot}. In turn, by virtue of Eqs.~\eqref{eq:SI} and~\eqref{eq:Pnet}, the photon-mediated heat $\dot{Q}_{\gamma}$ radiated between the two superconductors would be much smaller than what we found in our SQUID geometry (Fig.~\ref{fig6}).

Finally, let us consider the limiting case in which $S_1$ and $S_2$ are normal metals, i.e., resistors, connected by purely reactive (i.e., $\omega$-independent) superconducting lines, instead of Josephson junctions. In this situation the Josephson inductances $L_{J,\alpha}$ ($\alpha=1,2$) would vanish, and thus there would not be any impedance associated to the superconducting lines. The total impedance of the circuit would then simply be $Z_{\mathrm{tot}}=Z_1+Z_2$ (with $Z_i$ now real), and the final expression for the net radiated heat from 2 into 1 would reduce to the known result~\cite{Cleland2004,Meschke2006}:
\begin{align}
\dot{Q}_{\gamma}&=\frac{4Z_1 Z_2}{(Z_1+Z_2)^2} \int_0^\infty\,\frac{\mathrm{d}\omega}{2\pi}\,\hbar\omega\,[n_2(\hbar\omega)-n_1(\hbar\omega)]\cr
&=\frac{4Z_1 Z_2}{(Z_1+Z_2)^2}\,\frac{\pi k_B^2}{12 \hbar}\,\left(T_2^2-T_1^2\right).
\end{align}

\section{Conclusions}
\label{sec:conclusions}
In this work we have investigated the photon-mediated heat transport in Josephson-coupled BCS superconductors.
We have analyzed the photonic thermal conductance of a temperature-biased superconducting quantum interference device (SQUID), and found it to be several orders of magnitude smaller than the universal quantum of thermal conductance, $\kappa_0(T)=\pi k_B^2T/6\hbar$.
Our results are robust over a wide range of device parameters, such as the capacitance and critical current of the Josephson junctions, the temperature, and the asymmetry parameter between the junctions.
In particular, we have observed that increasing $C$, $I_c$ and $T$ enhances the coupling between the superconductors, thus resulting in a larger photon-mediated heat flow, whereas increasing the asymmetry $r$ between the junctions reduces the modulation contrast of the heat conductance.
For a temperature-biased SQUID with realistic setup parameters, we have shown that at low temperature the direct galvanic heat flow between the two superconductors exceeds by more than two orders of magnitude the photon-mediated one.
Nevertheless, the latter is typically much larger than the heat current exchanged between electrons and lattice phonons.
This shows that in principle by an appropriate nanodevice design one could make more relevant the photonics corrections. In fact, these  have a long range nature being connected to electromagnetic radiation while the galvanic (quasiparticle) effects which are more localised since connected to quasiparticle heat diffusion.

%
\section{Acknowledgments}
Fruitful discussions with C. Altimiras and T. T. Heikkil\"a are gratefully acknowledged.
R.B. and P.S. have been supported by MIUR-FIRB2013 - Project Coca (Grant No. RBFR1379UX).  P.S. and A.B. acknowledge funding from the European Union FP7/2007-2013 under REA Grant agreement No. 630925 - COHEAT and the COST Action MP1209. A.B. acknowledges MIUR-FIRB2012 - Project HybridNanoDev (Grant No.RBFR1236VV) and STM 2015 from CNR. 
F.G. acknowledges the European Research Council under the European Union's Seventh Framework Program (FP7/2007-2013)/ERC Grant agreement No.~615187-COMANCHE for partial financial support.

\end{document}